\documentclass[aps,prb,twocolumn,showpacs,superscriptaddress]{revtex4}

\usepackage[pdftex]{graphicx}
\usepackage[pdftex]{epsfig}
\usepackage{epstopdf}
\usepackage{color}
\usepackage{amssymb,amsmath}
\usepackage{graphicx}
\usepackage{dcolumn}
\usepackage{multirow}
\newcommand{\red}[1]{\textcolor{black}{#1}}

\begin{document}

\title{Tunneling in Nanoscale Devices}

\author{Mark Friesen}
\affiliation{Department of Physics, University of Wisconsin-Madison, Madison, WI 53706, USA}
\author{M. Y. Simmons}
\affiliation{Centre of Excellence for Quantum Computation and Communication Technology, School of Physics, University of New South Wales, Sydney NSW 2052, Australia}
\author{M. A. Eriksson}
\affiliation{Department of Physics, University of Wisconsin-Madison, Madison, WI 53706, USA}



\begin{abstract}
Theoretical treatments of tunneling in electronic devices are often based on one-dimensional (1D) approximations.
Here we show that for many nanoscale devices, such as widely studied semiconductor gate-defined quantum dots, 1D approximations yield an incorrect functional dependence on the tunneling parameters (e.g., lead width and barrier length) and an incorrect magnitude for the transport conductance.
Remarkably, the physics of tunneling in 2D or 3D also yields transport behavior that appears classical (like Ohm's law), even deep in the quantum regime.
\end{abstract}

\maketitle

Quantum effects such as tunneling play an expanding role in modern electronics.  
For many devices, including quantum dots and semiconductor qubits, tunneling processes are essential to device operation.\cite{VanDerWiel03,Hanson07,Zwanenburg13}
In dots, for example, tunneling mediates changes in the dot occupations, both real and virtual, while in qubits it plays a crucial role in gate operations.\cite{Elzerman04,Petta05}
A detailed knowledge of the tunnel rates can also improve the accuracy of qubit initialization and readout\cite{Hanson05} through a mechanism known as energy-dependent tunneling.\cite{MacLean07,Amasha08,Simmons10,Simmons11,Thalakulam11}

From a theoretical perspective, tunneling has long been perceived as a fundamentally one-dimensional (1D) process;\cite{Sethna82}
the modern formulation of quantum transport in terms of 1D channels embraces this view.\cite{Landauer87,Buttiker88}
It is therefore tempting to adopt a 1D approach when calculating transmission coefficients (e.g., along the ``path of least resistance"~\cite{Banks73,Banks73b}).
This approach may be effective for geometries that are intrinsically 1D, such as resonant tunneling diodes\cite{Klimeck95} and superconducting tunnel junctions.\cite{Griffin71}
However, tunnel rates in quantum dots exhibit considerable variability for small changes in device parameters, such as gate voltages.\cite{Thalakulam11,Schleser05}
Such behavior is still poorly understood, but it clearly depends on the structure of the tunnel barrier and the tails of the wave functions outside the nominal 1D transport channel.
In this paper, we show that 1D approximations yield very poor results when they are used to calculate transport coefficients in intrinsically 2D tunneling geometries.
We find that transverse spreading of the wave function suppresses the tunnel conductance, and that this effect is most prominent for devices with narrow leads and long tunnel barriers.

From a computational perspective, there are many technical challenges for modeling tunneling devices whose dimensions can range from nanometers to microns.\cite{Vasileska08}
Effective mass techniques are well suited for mesoscopic systems and can also be incorporated into multiscale simulations.\cite{derMaur11}
The conventional approach used in tunneling calculations involves scattering plane waves at a 1D barrier.\cite{DaviesBook,Hansen04}
This method employs open boundary conditions, which may be inappropriate, however, for closed systems like quantum dots.
Here, to obtain accurate estimates of transmission coefficients in nanoscale devices, we employ a new approach to modeling tunnel juntions based on closed boundary conditions.
We adopt an effective mass approximation and perform numerical simulations in simple but realistic device geometries.

\section{Tunneling, with closed boundary conditions} \label{sec:1}
Either open or closed boundary conditions can be used to model transport processes in nanoscale devices;\cite{Datta00} the appropriate choice is determined by the device geometry.
Over the past 20 years, an industry has grown around quantum transport calculations based on non-equilibrium Green's functions.
Typically, this method assumes open boundary conditions and a plane wave basis for the tunneling states.\cite{FerryBook}

Open boundary conditions are not necessarily suitable for quantum information applications.
For example, tunneling processes can mediate exchange interactions between static electron qubits in quantum dots;\cite{Loss98,Kane98}
in this scenario, the tunneling is virtual and closed boundary conditions are the most physical.
A complementary industry has grown around describing the evolution of semi-closed systems of qubits, based on master equations.\cite{NielsenBook}
However in closed systems, a rigorous formalism for computing the tunneling parameters (e.g., the transmission coefficients) has not yet been developed.

Here, we develop a theoretical model for tunneling in closed geometries.
We focus on the materials system of an electron gas confined to a semiconductor quantum well, where the electron wave functions can be treated as 2D due to strong quantum confinement in one direction.\cite{DaviesBook}
Although the electric field lines are 3D, we assume the lateral, 2D confinement potential can be shaped arbitrarily by voltages applied to electrostatic top-gates.
It is therefore sufficient to treat the one-electron tunneling problem as 2D.
The general methods developed here can also be applied to devices formed in $\delta$-doped layers.\cite{Feuchsle10,Feuchsle12}
The tunneling is 3D in that case, since there is no quantum well.
In the present work, we only consider 2D geometries.

The computational methods developed here are appropriate for tunneling between quantum dots, but they can also be extended to quasi-open systems such as leads, which we focus on in this work.
The theory is particularly well suited for numerical techniques such as finite element, effective mass simulations or tight binding methods with closed boundary conditions.
Here, we apply our theory to the problem of tunneling through 2D barriers with variable aspect ratios, using an effective mass approach.

Our focus is on the physics of \red{transmission coefficients rather than lead geometries.
(The latter impact tunneling mainly through the local density of states.\cite{Feuchsle10,Konemann01,Zumbuhl04,Fasth07,Lim09,Pierre10})
We therefore adopt a geometrical construction known as ``perfect" or ``ideal" leads,}
which are defined as lead regions far from a tunnel barrier, where the confinement potential is uniform in the longitudinal direction ($\hat{\bf y}$).\cite{DattaBook}
The corresponding momentum $k_y$ is a good quantum number.
We examine simple tunnel geometries formed of two perfect leads coupled through a square barrier.
We limit our attention to symmetric geometries where the subband energies are the same in both leads.

\begin{figure}[t]
\includegraphics[width=3in]{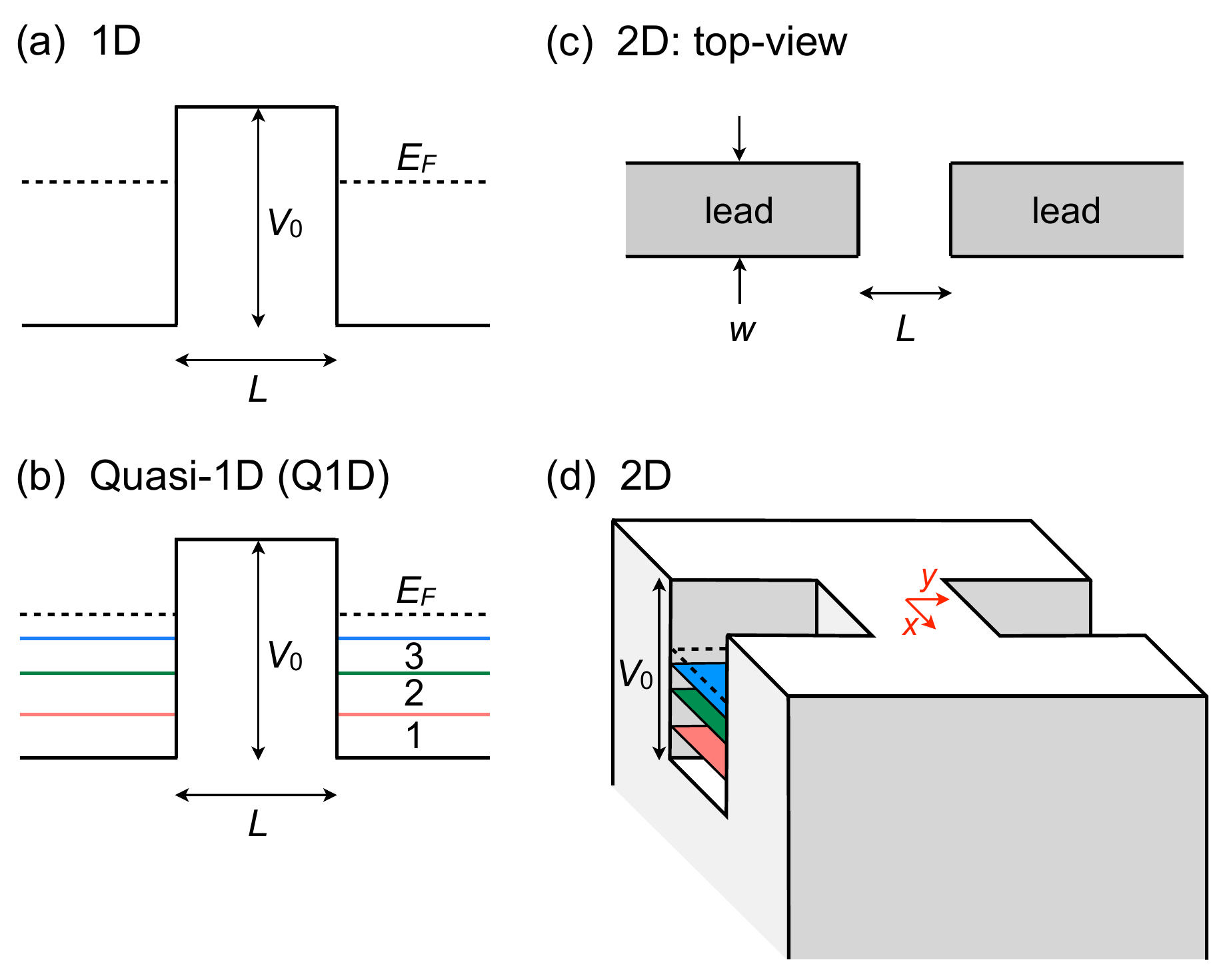}
\caption{\label{fig:models}
The quantum tunneling geometries considered in this work.
(a) A 1D model of a rectangular barrier, where tunneling occurs at the Fermi energy, $E_F$.
(b) A quasi-1D (Q1D) model, where band filling begins at the subband energies $E_{0,1}$-$E_{0,3}$, computed for infinite 2D leads.
(c) Top view of the 2D leads, with lead width $w$ and barrier length $L$.
(d) A 2D energy diagram of the leads.
The energy levels shown in the leads (from bottom to top) correspond to the first three subbands and the Fermi energy (dashed line).}
\end{figure}

Numerical simulations yield nearly-degenerate pairs of eigenstates, which are used to compute the transmission coefficients $t_\alpha$ for subband $\alpha$, following the methods developed in Appendices~\ref{app:A} and \ref{app:C}.
These molecular states are characterized as symmetric or antisymmetric, depending on whether the wave function has a node in the barrier.
We work at zero temperature, in the linear response regime, where tunneling occurs at the Fermi energy.
The total conductance is then expressed as a sum over the subbands:\cite{Landauer87,Buttiker88}
\begin{equation}
G=\frac{e^2}{h}\sum_{\alpha}g_{\alpha}|t_{\alpha }|^2 . \label{eq:LBmain}
\end{equation}
In this work, we consider a subband degeneracy of $g_\alpha=g_vg_s=4$, as appropriate for a 2D electron gas in Si, with a valley degeneracy of $g_v=2$, and a spin degeneracy of $g_s=2$.\cite{Friesen2007}

\red{The subband labels in Eq.~(\ref{eq:LBmain}) are defined far away from the barrier, in the perfect lead region, where the subbands are well defined.
In a conventional mode-matching analysis, where the scattering states are expanded in terms of ideal, 1D subbands and plane waves, the sum in Eq.~(\ref{eq:LBmain}) would extend over a vast number of states, and would include off-diagonal terms $t_{\alpha,\beta}$ between different subbands on the left- and right-hand sides of the barrier, due to the 2D nature of the perturbation (the tunnel coupling).
The theoretical method for computing transmission coefficients developed in Appendix~\ref{app:A}, combined with full 2D modeling, provides a highly efficient, 2D basis set  $\{\alpha\}$, which automatically includes the 2D perturbations.
In this method, the subband labels are still well defined, because the modes adiabatically extend to the perfect leads.
The 2D basis is analogous to the highly efficient bases that have been used in configuration interaction calculations of multi-electron quantum dots.\cite{Friesen2003,Nielsen2012}
For the symmetric barrier geometries considered here, a given mode $\alpha$ on the left-hand side of the barrier is always matched to the same mode on the right-hand side of the barrier, in a symmetric or antisymmetric configuration, yielding transmission coefficients in Eq.~(\ref{eq:LBmain}) that are purely diagonal:  $t_\alpha \equiv t_{\alpha,\alpha}$.
This is not true for asymmetric geometries.}

\section{Lead geometries}
We now characterize the dependence of tunneling on the tunnel barrier dimensionality by comparing analogous 1D and 2D lead geometries.
We consider the 1D geometry shown in Fig.~\ref{fig:models}(a), consisting of a finite barrier of height $V_0$ between two finite, symmetric leads.
(The infinite barriers at the outer edges of the leads are not shown.)
We also consider the 2D geometry shown in Fig.~\ref{fig:models}(d), which has the same longitudinal cross-section as the 1D geometry.
In the transverse direction, we treat the leads as square wells with finite width $w$, and a surrounding confinement barrier of height $V_0$ -- the same as the tunnel barrier.
\red{Although realistic lead geometries are not typically as regular or as rectangular as those indicated in Figs.~\ref{fig:models}(c) and (d), we note that only the shape of the \emph{tip} of the lead plays a prominant role in determining the tunneling transmission.\cite{Bonnell2001}
The simulation geometries considered in Sec.~\ref{sec:results} cover a range of tip widths $w$, barrier lengths $L$, and barrier heights $V_0$ of interest in nanoscale tunneling devices, such as quantum dots. 
In particular, this range includes the narrow constrictions and low barriers associated with depletion gates in qubit devices.}

We cannot compare the 1D and 2D geometries directly, since the latter contains subbands, while the former does not.
We therefore consider a third, \emph{quasi}-1D model (Q1D), sketched in Fig.~\ref{fig:models}(b), which incorporates subbands in the following manner.
First, we note that the 1D and 2D models both describe bands of electrons that fill up to the Fermi energy, $E_F$. 
The lowest mode in the 1D model has zero energy, while the lowest mode in subband $\alpha$ of the 2D model has energy $E_{0,\alpha}>0$, due to the transverse confinement.
The Q1D model is obtained by defining a new 1D model for each subband, where the minimum potential energy (inside the lead) is renormalized to the value $E_{0,\alpha}$. 
This model yields the correct filling for each subband.
The total conductance is the sum of the contributions from the effective subbands.
An important physical distinction between the 2D and Q1D models is that 2D wave functions spread out transversely into the tunnel barrier, while Q1D wave functions do not.

\begin{figure}[t]
\includegraphics[width=3.3in]{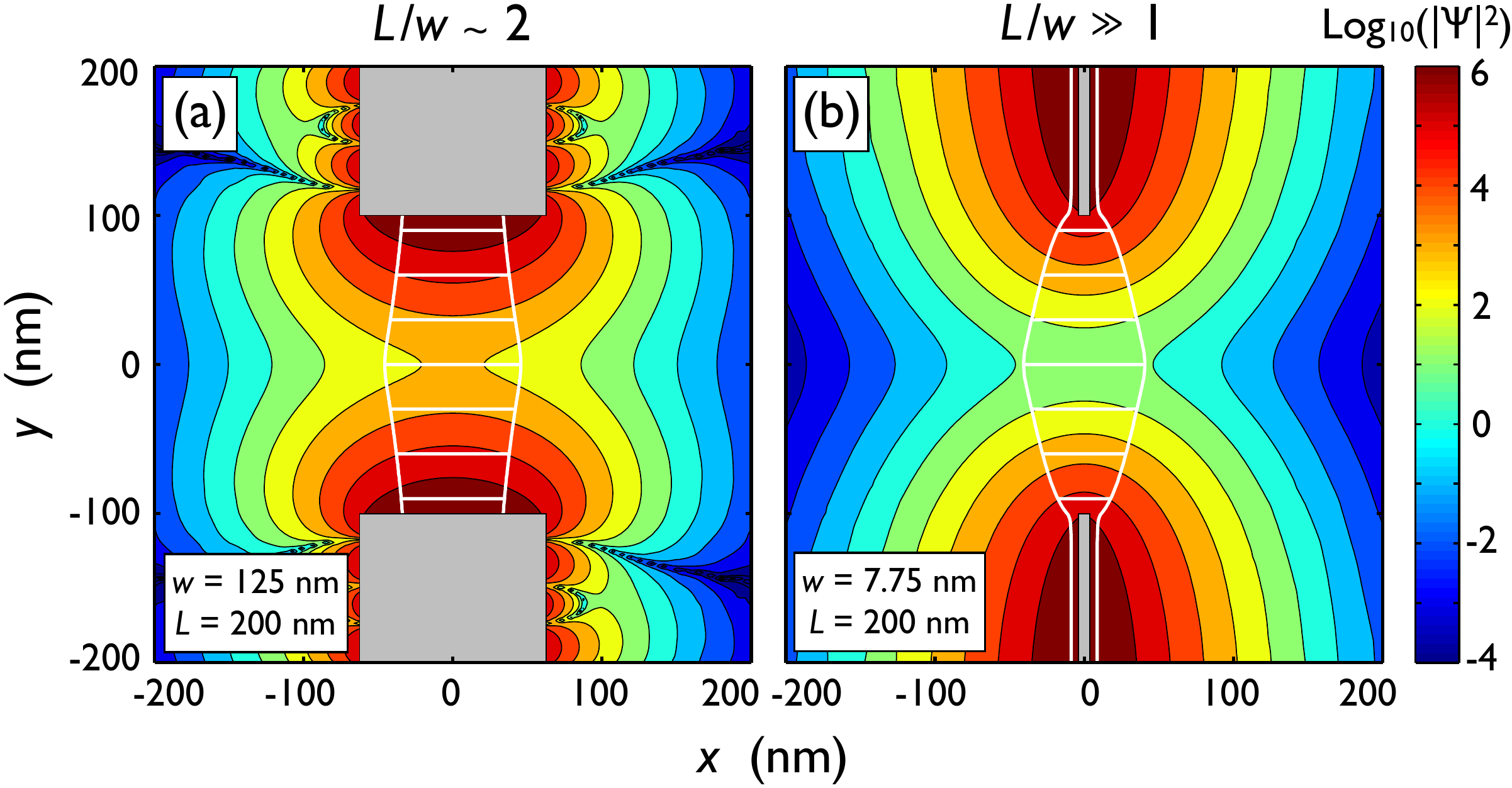}
\caption{\label{fig:FWHM}
Electron probability densities in two devices with the same, long tunnel barrier ($L=200$~nm), the same Fermi energy ($E_F=2.5$~meV), and same barrier height ($V_0=3$~meV, or 0.5~meV above the Fermi energy).
(a) Wide-lead geometry ($w=125$~nm).
(b) Narrow-lead geometry ($w=7.75$~nm).
Contours correspond to factors of 10 in the probability density.
Horizontal white lines show constant-$y$ ``cuts" through the probability density.
The length of the cut indicates the full-width-at-half-max (FWHM) of the peak in density.
Vertical white curves indicate the FWHM as a function of position ($y$).
In (a), the FWHM is always narrower than $w$.
In (b), the FWHM is always wider than $w$ (becoming much wider inside the tunnel barrier), indicating a transverse spreading of the wave function. }
\end{figure}

\section{Results} \label{sec:results}
The tunnel barriers in electrostatically-defined nanoscale devices such as qubits are typically low.
\red{Energy-dependent tunneling experiments suggest barrier heights of order meV.\cite{MacLean07,Amasha08,Simmons10,Simmons11,Thalakulam11}}
Here, we perform simulations on geometries with barriers in the range 0.5-10~meV (measured from the Fermi energy).
We use the transverse effective mass appropriate for Si, $m^*=0.19\, m_e$, and the Fermi energy $E_F=2.5$~meV, corresponding to a typical electron sheet density of $n\simeq 4\times 10^{11}$~cm$^{-2}$.

\begin{figure}[t]
\includegraphics[width=3.3in]{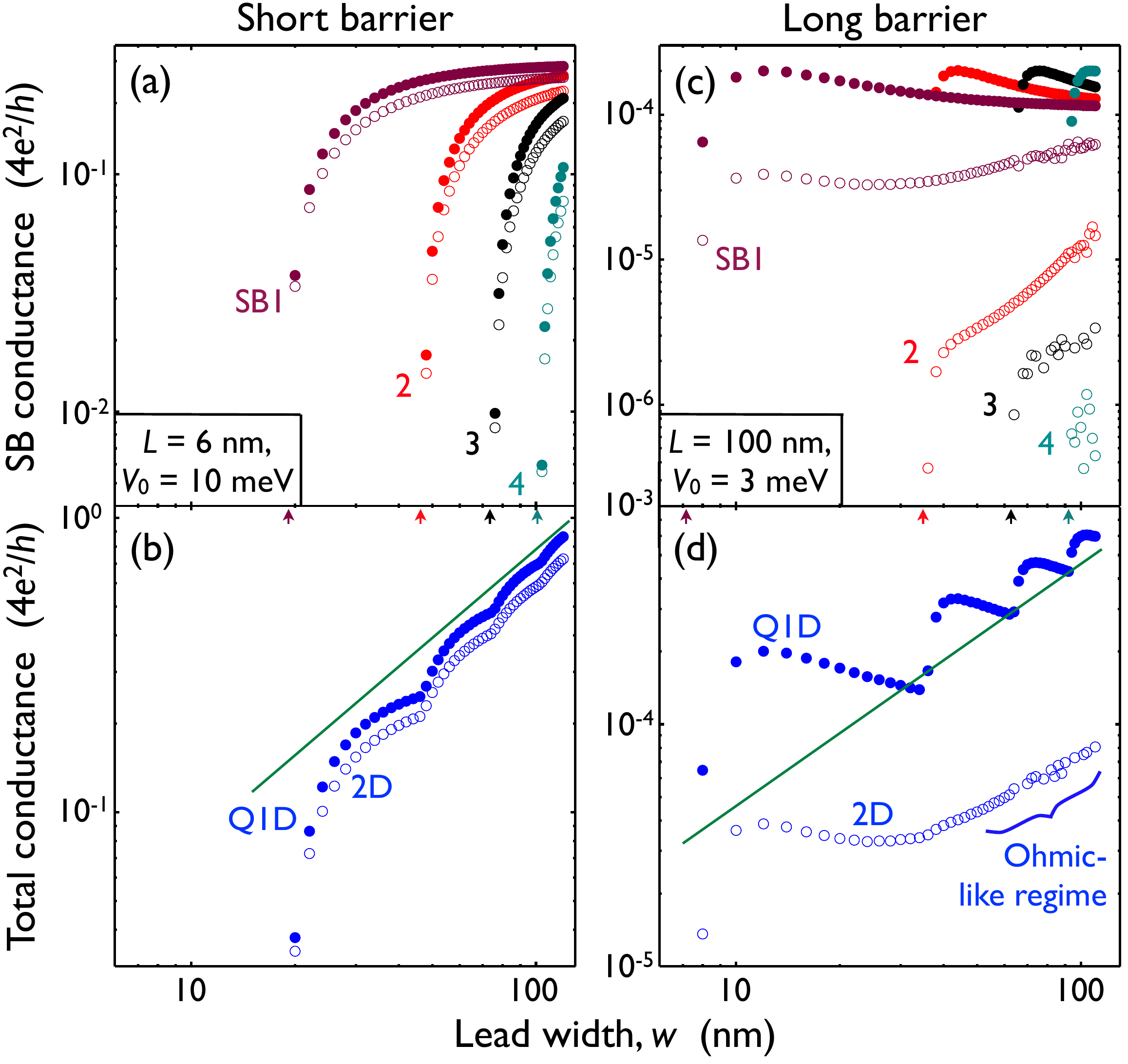}
\caption{\label{fig:Gvswsame}
Conductances in short and long-barrier devices, calculated for a Q1D model (filled circles) and a 2D model (open circles).
The Fermi energy is the same for all calculations ($E_F=2.5$~meV).
The barrier heights ($V_0$) and barrier lengths ($L$) are indicated.
The two models agree quite well for the short-barrier geometry.
For the long-barrier geometry, the models disagree in both form and magnitude.
(a) and (b) have the same values of $V_0$ and $L$.
(a) shows the contributions from each of four subbands (SB 1-4) as they begin to fill for increasing lead widths.
(b) shows the the total conductance (the sum of the subband contributions), for both the Q1D and 2D models.
The onset of subband filling is indicated with small arrows.
The solid green line corresponds to the asymptotic Q1D conductance derived in Appendix~\ref{app:D}. 
(c) and (d) are analogous to (a) and (b), but with different values of $L$ and $V_0$.
In (d), the 2D conductance is much smaller and smoother than the Q1D estimate.}
\end{figure}

In Figs.~\ref{fig:FWHM} and \ref{fig:Gvswsame}, we present several studies with fixed barrier height $V_0$ but variable lead width $w$.
Electron probability densities are shown in Figs.~\ref{fig:FWHM}(a) and (b) for devices with relatively long tunnel barriers.
Both solutions are obtained at the Fermi energy.
While several subbands are filled at this energy [particularly for the wide lead geometry of Fig.~\ref{fig:FWHM}(a)], we only present results for the lowest subband.
Several transverse cuts are highlighted in the barrier regions, with the length of the cut indicating the full-width-at-half-max (FWHM) of the probability density.
For the wide-lead geometry in Fig.~\ref{fig:FWHM}(a), the FWHM changes very little along $y$ (as consistent with a 1D tunneling model), and is always narrower than the lead width, $w$.
For the narrow leads in Fig.~\ref{fig:FWHM}(b), the FWHM balloons out to more than ten times the lead width; such behavior cannot be accurately represented in a 1D model.
The transverse spreading can be viewed as a geometrical effect that suppresses the wave function, and ultimately the tunnel conductance.

An additional suppression of the wave function in the tunnel barrier arises from the strong transverse confinement in the leads. 
This can also be observed in Fig.~\ref{fig:FWHM}, where for the same total energy, we can see the different relative contributions from the transverse and longitudinal energy quadratures for the two different lead widths. 
The weaker confinement in the wide-lead geometry yields a larger longitudinal momentum, as evidenced by the presence of nodes along the edges of the lead in Fig.~\ref{fig:FWHM}(a).
The resulting probability density in the tunnel barrier is about 100 times greater than  Fig.~\ref{fig:FWHM}(b), 
despite the barriers having the same length and height. 
This effect has two consequences for our tunneling simulations: 
(1) at the onset of filling in a given subband, the wave function (and the conductance) are strongly suppressed because \emph{all} the energy is associated with transverse confinement; 
(2) when several subbands are filled, the conductance contributions from higher subbands are suppressed, because their longitudinal momentum is smaller.

Combined, these two effects yield a suppression of the tunnel conductance in devices with narrow leads and/or long tunnel barriers. 
This is evident in Fig.~\ref{fig:Gvswsame}, where conductances are plotted for devices with short barriers [Figs.~\ref{fig:Gvswsame}(a) and (b)] vs.\ long barriers [Figs.~\ref{fig:Gvswsame}(c) and (d)].
For the devices in Figs.~\ref{fig:Gvswsame}(a) and (b), there is little transverse spreading of the wave function in the tunnel barrier, yielding Q1D and 2D conductance estimates that are nearly equal.
For such short-barrier devices, the Q1D model provides an accurate description of the tunneling.

Figures~\ref{fig:Gvswsame}(c) and (d) correspond to long-barrier devices that are common in qubit experiments.
For this geometry, the true 2D conductances are strongly suppressed relative to the Q1D model.
For several of the subbands in Fig.~\ref{fig:Gvswsame}(c), the suppression factor is over two orders of magnitude.
\red{It is interesting to note that a strong modulation of the conductance has also been observed in the ballistic transport regime ($E>V_0$),\cite{Yacoby1994} which can similarly be attributed to the breakdown of 1D transport models.\cite{Jauho1999}}

Figures~\ref{fig:Gvswsame}(a) and (b) indicate that tunneling is effectively 1D for devices with short barriers and very low aspect ratios ($L/w\ll 1$).
It is interesting to ask whether this behavior persists for devices with longer barriers.
To answer this question, we have obtained an analytical expression for the Q1D conductance in the asymptotic (large-$w$) limit of the Q1D model in Appendix~\ref{app:D}, which we plot as solid lines in Figs.~\ref{fig:Gvswsame}(b) and (d).
In Fig.~\ref{fig:Gvswsame}(b) the Q1D and 2D results both approach the Q1D asymptote at large $w$.
For the long-barrier devices in Fig.~\ref{fig:Gvswsame}(d), the 2D conductance does not approach the Q1D asymptote for lead widths $w$ consistent with typical quantum dot devices. 
Indeed, the total 2D conductance in Fig.~\ref{fig:Gvswsame}(d) is suppressed by about an order of magnitude compared to the Q1D asymptote.

\section{Ohmic Behavior}
Perhaps the most remarkable feature of the 2D conductance in Fig.~\ref{fig:Gvswsame}(d) is the emergence of Ohmic-like behavior, as characterized by the smooth, linear dependence $G\propto w$.
(Classically, the resistance varies as $R=R_sL/w$ for a 2D device with sheet resistance $R_s$ and length $L$.
\red{In Fig.~\ref{fig:Gvswsame}(d), we note that the slope of the 2D conductance is slightly less than 1; however it is expected to approach 1 in the asymptotic, large-$w$ limit where the Q1D and 2D models are equivalent.})
This is in stark contrast with the Q1D model, for which the conductance exhibits quantum steps at the onset of filling in each subband (indicated by small arrows in the figure).
The Q1D conductance asymptotes (plotted as solid lines) have a similar, smooth appearance.
However, the onset of classical, Ohmic behavior in the 2D model is already prominent for very narrow leads; it emerges long before the quantum steps disappear in the Q1D model.
We therefore conclude that the transverse spreading of the wave function has a strong effect on the functional form of the conductance, in addition to suppressing its magnitude, which cannot be explained by the Q1D tunneling model.

\section{Summary and Conclusions}
We have developed a technique for analyzing tunneling in two or three dimensions and applied it to nanoscale tunneling geometries in a two-dimensional electron gas in Si.
The scheme employs closed boundary conditions, making it ideal for finite element or finite difference methods.
We find that approximate, Q1D models of tunneling in devices of experimental interest yield poor results because the real 2D wave functions spread out transversely in the tunnel barrier, causing a strong suppression of the conductance.
Such fundamentally 2D behavior is most evident in devices with narrow leads and long tunnel barriers.
In cases where transverse spreading has a strong effect on the magnitude of the tunnel conductance, it also affects the functional form by smoothing out the quantum steps associated with subband filling, and mimicking classical, Ohmic-type behavior.
Since the leads formed by depletion gates in a two-dimensional electron gas tend to be 
\red{constricted} and narrow in typical nanoscale devices, we expect Q1D models to provide a rather poor description of the tunneling in these systems.

\begin{acknowledgements}
We are grateful to S. N. Coppersmith, J. K. Gamble, B. Koiller, W. Pok,  and A. Saraiva for stimulating discussions. 
This work was supported by National Science Foundation (DMR-0805045, DMR-1206915), the United States Department of Defense, and the Australian Research Council Centre of Excellence for Quantum Computation and Communication Technology (CE110001027).  M.\ Y.\ S.\ acknowledges an ARC Federation Fellowship.
The US government requires publication of the following disclaimer:  the views and conclusions
contained in this document are those of the authors and should not be
interpreted as representing the official policies, either expressly or
implied, of the US Government.  
\end{acknowledgements}

\appendix


\section{Tunneling with Closed Boundary Conditions} \label{app:A}
The goal of our tunneling calculations is to compute the transmission coefficients $t_{\alpha,\beta}$ from subband $\alpha$ in the left lead to subband $\beta$ in the right lead, across the symmetric, 2D tunnel barrier $V(x,y)=V(x,-y)$.  
(The theory can trivially be extended to 3D.)
The Landauer-B\"{u}ttiker formula for linear conductance is given by 
\begin{equation}
G=\frac{e^2}{h}\sum_{\alpha,\beta}g_{\alpha,\beta}|t_{\alpha,\beta}|^2 , \label{eq:LBfull}
\end{equation}
where $g_{\alpha,\beta}$ is the degeneracy of the transport channel.
For very low temperatures, the sum extends over all the subbands in the lead whose energies lie below the Fermi energy.

Equation~(\ref{eq:LBfull}) is normally derived in terms of momentum states, which are the eigenstates inside regions known as ``perfect leads," and characterized by $\partial V/\partial y=0$.
Here, $\hat{\mathbf{y}}$ is the longitudinal direction for tunneling, as indicated in Fig.~1 of the main text.
The leads are confined in the transverse direction, and the transverse eigenstates are called subbands.
(They form energy bands due to the continuum of longitudinal modes.)
In the barrier region, where $\partial V/\partial y \neq 0$, the tunnel barrier can be viewed as a 
\red{2D perturbation that couples modes in this basis set of ideal, 1D subbands and longitudinal} momentum states.

\begin{figure}[t]
\includegraphics[width=2.2in]{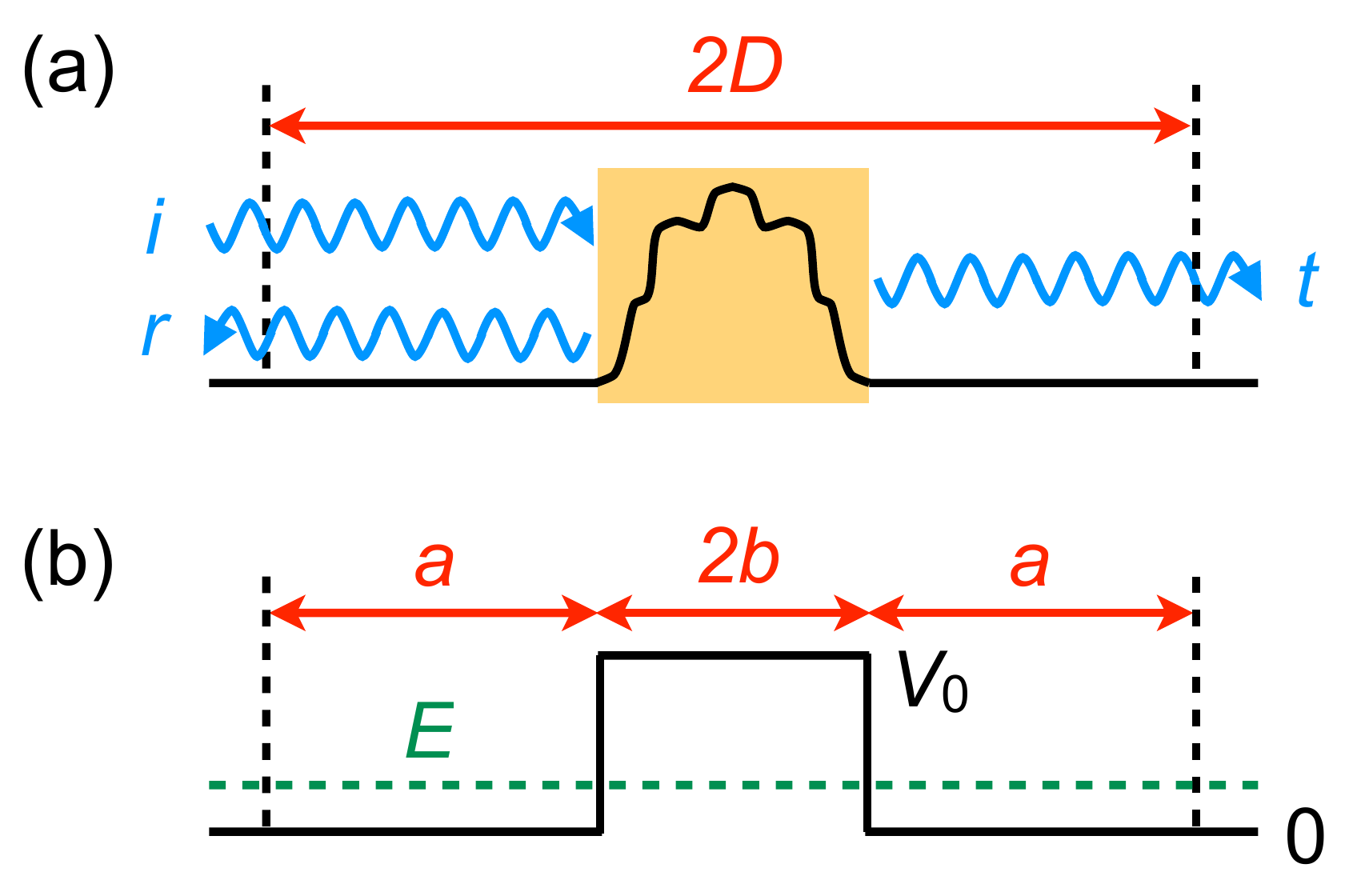}
\caption{\label{fig:bar1D}
(a) A 1D tunneling geometry with an arbitrary barrier shape.
The incident, reflected and transmitted scattering states are shown,  
and left-right mirror symmetry is assumed.
(b)  A square barrier geometry, with barrier height $V_0$.  
The tunneling state with energy $E$ is assumed to lie at the Fermi energy, $E_F$, for the case of zero bias and zero temperature.
The vertical dashed lines indicate infinite barriers that are introduced for the case of closed boundary conditions.}
\end{figure}

We begin by considering a 1D tunnel barrier of arbitrary shape, but with mirror symmetry, as shown in Fig.~\ref{fig:bar1D}.
The formalism we develop can easily be extended to a more general geometry.
For the symmetric case, however, the momentum labels are the same on the left- and right-hand sides of the barrier.
For 1D geometries, there are no subbands, and we can drop the $\alpha$ label.
We then focus on the diagonal coefficients $t_\alpha\equiv t_{\alpha,\alpha}(k)$, which do not involve subband mixing. 
In this case, we have
\begin{equation}
G\simeq \frac{e^2}{h}\sum_{\alpha}g_{\alpha}|t_{\alpha}|^2 . \label{eq:LB}
\end{equation}
\red{For 2D geometries, the 2D simulations described in Appendix~\ref{app:C} yield solutions that automatically include the couplings between ideal, 1D subbands, as discussed in Sec.~\ref{sec:1}.
For asymmetric geometries, a given subband $\alpha$ on the left-hand side of the barrier can scatter to a different subband $\beta$ on the right-hand side.
In this case, one should use Eq.~(\ref{eq:LBfull}).
For symmetric barriers, however, symmetry only allows diagonal matching of the 2D modes, in symmetric and antisymmatric combinations.
In this case, as for the 1D symmetric barrier, Eq.~(\ref{eq:LB}) is appropriate.}

Inside the perfect leads, the wave function is a plane wave with momentum $k$ determined by the energy $E$, and a phase that is determined by the details of the barrier.
In a perturbative scattering theory, our calculation therefore reduces to calculating this one phase parameter, regardless of the complexity of the barrier.

We now focus on closed geometries, as discussed in the main text.
In the transfer matrix approach, the matrix $\mathcal{T}$ relates the scattering amplitudes $a_{L,R,\pm}$ between plane waves with positive $(+)$ and negative $(-)$ momenta, on the left- $(L)$ and right-hand-sides $(R)$ of the barrier as
\begin{equation}
\mathcal{T} \begin{pmatrix} a_{L,+} \\ a_{L,-} \end{pmatrix}
= \begin{pmatrix} a_{R,+} \\ a_{R,-} \end{pmatrix} .  \label{eq:T1}
\end{equation}
Because of the imposed symmetry constraint, the scattering states on the left- and right-hand-sides of Fig.~\ref{fig:bar1D}(a), have the same absolute value of the momentum, $k=k_L=k_R$.

The transfer matrix in Eq.~(\ref{eq:T1}) is conventionally solved by considering an incident plane wave with momentum $+k$ and unit amplitude, $a_{L,+}=1$, a reflected wave with momentum $-k$ and amplitude $a_{L,-}=r$, and a transmitted wave with momentum $+k$ and amplitude $a_{R,+}=t$.
In this arrangement, $a_{R,-}=0$.
The time-reversed solution involves states with the same left- and right-going momenta, so the same transfer matrix should describe both situations,\cite{DaviesBook} such that $a_{L,+}=r^*$, 
\red{$a_{L,-}=1$}, 
$a_{R,+}=0$, and $a_{R,-}=t^*$.
Adopting the normalization condition $|r|^2+|t|^2=1$, the transfer matrix can be uniquely expressed in terms of the reflection and transmission coefficients:
\begin{equation}
\mathcal{T} = 
\begin{pmatrix} 1/t^* && -r^*/t^* \\ -r/t && 1/t \end{pmatrix} 
. \label{eq:Msolve}
\end{equation}
Recall that the transmission and reflection coefficients are implicit functions of the momentum.
Explicit knowledge of the barrier allows us to compute this dependence.

As written above, Eq.~(\ref{eq:Msolve}) is usually interpreted as the forward momentum transfer between leads with open boundary conditions.\cite{DaviesBook,FerryBook}
More generally, it describes the coherent coupling between leads on either side of the barrier.  
We can make use of this fact to investigate \emph{stationary} solutions, with closed boundary conditions, such as the geometry within the infinite dashed barriers indicated at positions $y=\pm D=\pm (a+b)$ in Figs.~\ref{fig:bar1D}(a) and (b).
In the left and right perfect lead regions of Fig.~\ref{fig:bar1D}(a) (far from the tunnel barrier), the bound state solutions are given by
\begin{equation}
\psi_L=\sin [(y+D)k]  \,\,\,\,\,\, \text{ and } \,\,\,\,\,\,  \psi_R=\pm \sin [(y-D)k] , \label{eq:bound}
\end{equation}
where we have neglected the normalization.
Again, we note that these solutions do not depend on the shape of the barrier, only on the existence of perfect leads.
(The wave function has a more complicated behavior near the barrier, of course.
However, we do not solve for the wave function in this region.)
The $+$  and $-$ signs on the right-hand-side of Eq.~(\ref{eq:bound}) correspond to symmetric $(S)$ and antisymmetric $(A)$ solutions, respectively, with momenta $k_{S,A}$ and energies $E_{S,A}=\hbar^2 k_{S,A}^2/2m^*$.
These are also referred to as bonding and antibonding molecular orbitals.
Here, $m^*$ is the effective mass.  

Equation~(\ref{eq:Msolve}) is solved by noting that the reflection and transmission coefficients, $r$ and $t$, are regular functions of $k$, and that the eigenstates $k_S$ and $k_A$ occur in closely spaced pairs.
To a good approximation, we therefore expect the \emph{same} $r$ and $t$ to apply to both $k_S$ and $k_A$.
Expanding Eq.~(\ref{eq:bound}) in terms of plane waves, we then obtain
\begin{eqnarray}
|t|^2 &\simeq& \sin^2 \left[ (k_A-k_S)D \right]    \label{eq:sin2} \\
&=& \sin^2 \left[ \sqrt{2m^*D^2/\hbar^2}\left( \sqrt{E_A} -\sqrt{E_S} \right)\right] . \nonumber
\end{eqnarray}
\red{An equivalent result was previously obtained for a symmetric 1D tunnel barrier, in a different context.\cite{Sushkov2001}}

Equation (\ref{eq:sin2}) is quite general.
We have placed no constraints on the shape of the tunnel barrier, except for its symmetry.
In this formalism, further progress requires solving a Schr\"{o}dinger equation for the nearly degenerate energies $E_A$ and $E_S$.
Many efficient techniques are known for solving the  Schr\"{o}dinger equation with closed boundary conditions.
In the main text and below, we assume numerical solutions based on finite element methods.

For tunneling in dimensions greater than one, we must take into account the subband structure.
In this work, we focus on 2D geometries.
As indicated in Eq.~(\ref{eq:LB}), each filled subband acts as a transport channel that contributes to the tunneling conductance.  
The perfect leads allow us to assign energy labels to these subbands.\cite{DattaBook,Lent90}
For tunneling in higher dimensions, the perfect lead formalism also provides a convenient  means of distinguishing the transverse and longitudinal components of the momentum.
The former is associated with the subband confinement and plays no direct role in the tunneling.\cite{DaviesBook,Hansen04}
The momenta $k_A$ and $k_S$ appearing in Eq.~(\ref{eq:sin2}) refer only to the longitudinal degrees of freedom.

To separate the longitudinal and transverse components, we consider the region of the device far from the barrier, well inside the perfect leads.
In this local region, the 2D confinement potential is separable [since $V(x,y)=V(x)$], so the total energy is a linear combination of its transverse and longitudinal components.  
The transverse component is just the subband energy $E_{0,\alpha}$, which can be computed by solving the 1D Schr\"{o}dinger along a cut $V(x)$ in the transverse direction.
The total energy is obtained by solving the full 2D geometry, numerically, yielding nearly-degenerate pairs of eigenstates, $E_{A,\alpha}$ and $E_{S,\alpha}$, similar to the 1D case.
Equation~(\ref{eq:sin2}) is then straightforwardly modified to describe just the longitudinal components:
\begin{eqnarray}
|t_\alpha|^2 &\simeq& \sin^2 \left[ \sqrt{2m^*D^2/\hbar^2}\left( \sqrt{E_{A,\alpha}-E_{0,\alpha}}
\right. \right.  \nonumber  \\ && \left. \left. \hspace{.35in}
-\sqrt{E_{S,\alpha}-E_{0,\alpha}} \right)\right]. \label{eq:sin23DS}
\end{eqnarray}
When numerical methods are used, each eigenstate must be analyzed to determine its subband label $\alpha$.

\section{Errors Due to Finite-Size Leads} \label{app:B}
The error in Eq.~(\ref{eq:sin2}) arises from our assumption that the same reflection and transmission coefficients, $r$ and $t$, can be used to describe $k_S$ and $k_A$, despite the fact that $r$ and $t$ are functions of the momentum.
This error disappears in the limit of infinite leads ($D\rightarrow \infty$).
(To see this, we note that the fraction of the wavefunction probability inside the barrier region, where the $S$ and $A$ modes differ, goes to zero.  Hence $E_A\rightarrow E_S$ and $k_A\rightarrow k_S$.)
We may therefore view the error in Eq.~(\ref{eq:sin2}) as ``truncation error" arising from the finite-size leads.
In this Appendix, we provide an estimate of this truncation error.
We make use of analytical solutions that can be obtained for 1D square barriers, and we compare the results for open and closed boundary conditions.

We consider the 1D square barrier geometry shown in Fig.~\ref{fig:bar1D}(b), with a barrier of height $V_0$ extending from $y=-b$ to $b$.
In the leads, the momentum and energy of the plane wave tunneling eigenstates are related by $E=\hbar^2k^2/2m^*<V_0$.
In the barrier, the solutions decay with wave vector $q$ defined by $V_0-E=\hbar^2q^2/2m^*>0$.

For the open geometry shown in Fig.~\ref{fig:bar1D}(b), where the lead regions extend out to infinity, the  conventional scattering states in the left, middle, and right-hand regions are given by
\begin{eqnarray}
\psi_L &=& e^{iky}+r e^{-iky} , \label{eq:pL} \\
\psi_M &=& A e^{-qy}+ B e^{qy} , \\
\psi_R &=& t e^{iky} , \label{eq:pR} 
\end{eqnarray}
respectively.
Here and below, we ignore overall normalization.
By matching Eqs.~(\ref{eq:pL})-(\ref{eq:pR}) and their first derivatives at the boundaries, we can obtain an exact expression for the transmission coefficient:\cite{DaviesBook}
\begin{eqnarray}
|t|^2 &=& \frac{4k^2q^2}{4k^2q^2+(k^2+q^2)^2 \sinh^2 (2qb)} \nonumber \\
&& \hspace{.3in} 
\simeq \frac{16k^2q^2 e^{-4qb}}{(k^2+q^2)^2} , \label{eq:texact}
\end{eqnarray}
where the approximate result in the second line corresponds to the ``large-barrier limit," $\exp(-4qb)\ll 1$, for which the correction terms are exponentially small.
(Note that in the main text, we use the notation $L=2b$.)

The large-barrier limit is usually quite appropriate for nanoscale devices.
To take an example, we consider a particularly low quantum barrier of height $\sim 1$~meV and width $\sim 50$~nm, which is characteristic of single-electron tunneling.
Even in this case, the difference between the two expressions in Eq.~(\ref{eq:texact}) is $\sim 3\%$.
(Here, we assume a Si quantum well, with $m^* \simeq 0.2\,m_0$.)
For many other tunneling geometries of interest, the barrier is even larger, and the approximation in Eq.~(\ref{eq:texact}) is excellent.

For the closed geometry shown in Fig.~\ref{fig:bar1D}(b), with infinite barriers introduced at $y=\pm (a+b)$, the eigenstate solutions are given by
\begin{eqnarray}
\psi_L &=& \sin [(y+a+b)k] , \\
\psi_M &=& C(e^{-qy}\pm e^{qy}) , \\
\psi_R &=& \pm \sin [(y-a-b)k] ,
\end{eqnarray}
where the $\pm$ signs correspond to the symmetric $(k_S,q_S)$ or antisymmetric $(k_A,q_A)$ solutions, respectively.
By matching the wave functions and their derivatives at the boundary, we obtain exact solutions, expressed transcendentally as
\begin{gather}
\tan (k_Sa) \tanh (q_Sb) = -\frac{k_S}{q_S} , \label{eq:1DS} \\
\tan (k_Aa) = -\frac{k_A}{q_A}  \tanh (q_Ab) , \label{eq:1DA}
\end{gather}
along with the relations
\begin{equation}
\frac{\hbar^2}{2m^*}\left( k_{S,A}^2+q_{S,A}^2 \right) = V_0 . \label{eq:balance}
\end{equation}

We can use these results to obtain a simple, approximate expression for the transmission coefficient.
Defining $k_S=k$ and $k_A=k+dk$, we can expand Eqs.~(\ref{eq:1DS})-(\ref{eq:balance}) in terms of the small parameter $dk/k \ll 1$.
In the large-barrier limit, Eq.~(\ref{eq:sin2}) then yields
\begin{equation}
|t|^2 \simeq \frac{16k^2q^2 e^{-4qb}}{(k^2+q^2)^2}
\left[ \frac{1+(b/a)}{1+1/(qa)} \right]^2 . \label{eq:t1Dapprox}
\end{equation}

We can finally compare the open and closed solutions for a 1D square tunnel barrier.
For infinite leads ($a\rightarrow \infty$), the transmission coefficients in Eqs.~(\ref{eq:texact}) and (\ref{eq:t1Dapprox}) are identical.
At the next order of approximation, with $a \gg b,q^{-1}$, we obtain
\begin{equation}
\frac{|t_\text{closed}|^2}{|t_\text{open}|^2} 
\simeq 1+\frac{2(b-q^{-1})}{a} . \label{eq:trunc}
\end{equation}
For the 1D square barrier geometry, the leading order truncation error in $|t|^2$ is given by the second term on the right-hand side of Eq.~(\ref{eq:trunc}).
Note that $0<q^{-1}\lesssim 4b$ is satisfied for devices in the large-barrier limit, which we have argued includes most devices of interest.
We conclude that using long leads in our simulations ($a\gg b$) ensures accurate results in Eq.~(\ref{eq:sin2}).

\section{Numerical Methods} \label{app:C}
In this Appendix, we outline the numerical procedures used to solve for the transmission coefficients derived in Appendix~\ref{app:A}.  
The results of typical numerical calculations are reported in the main text.

For Q1D geometries, we assume that the subband energies $E_{0,\alpha}$ are known.
(They are solved separately, using a 1D Schr\"{o}dinger equation, as described in Appendix~\ref{app:D}.)
We then solve for $k_S$ and $q_S$, simultaneously, for a fixed value of $b$ in Eqs.~(\ref{eq:1DS}) and (\ref{eq:balance}).
Here, we replace $V_0\rightarrow \widetilde{V}_0=V_0-E_{0,\alpha}$ to account for the subband confinement in the Q1D model, as discussed in the main text.
The lead length $a$ is chosen to provide the desired energy $\hbar^2k_S^2/2m^*=\widetilde{E}_F$, where we have also substituted $E_F\rightarrow \widetilde{E}_F=E_F-E_{0,\alpha}$.
There are many such solutions for $a$ (they form a discrete set, corresponding to increments in $a$ that satisfy $\Delta a\simeq 2\pi/k_S$).
For numerical accuracy, we choose a value such that $a\gg b,q_S^{-1}$, as described in Sec.~B, above.
Once $a$ is known, we can solve Eqs.~(\ref{eq:1DA}) and (\ref{eq:balance}) simultaneously to determine $k_A$ and $q_A$.

We follow a similar procedure for 2D geometries.
We first determine the subband energies $E_{0,\alpha}$ by taking a transverse, 1D cut through the 2D confinement potential, deep in the perfect lead region.
We then solve for $E_{\alpha,S}$ and $E_{\alpha,A}$ using a numerical 2D Schr\"{o}dinger solver in the full 2D geometry.
As in the Q1D case, we choose an appropriate value of $a$, such that $E_{\alpha,S}=E_F$.
For both the Q1D and 2D models, we require our solutions to satisfy $a>11b$, for numerical accuracy.
For the results reported in the main text, we typically use $a\simeq 600$~nm.

\section{Asymptotic Conductance for the Q1D Model} \label{app:D}
In this Appendix, we obtain the exact asymptotic behavior of the tunnel conductance for a Q1D square barrier geometry, in the limit of wide leads.
Typical Q1D and 2D geometries are shown in Fig.~1 of the main text.

When the lead width $w$ is large but finite, a large number ($N_w$) of subbands contributes to the tunnel conductance:
\begin{equation}
G=\frac{4e^2}{h}\sum_{\alpha=1}^{N_w} |t_\alpha|^2 . \label{eq:GNw}
\end{equation}
In the limit $w\rightarrow \infty$, the set forms a continuum.  
The subband energies fall into the range $E_{0,\alpha}\in [0,E_F]$.  
We can therefore express the conductance as an integral:
\begin{equation}
G=\frac{4e^2}{h}\int_0^{E_F}\!\!dE\left(\frac{\partial N_w}{\partial E}\right) |t(E)|^2 . \label{eq:GNwint}
\end{equation}
We adopt the large-barrier limit in Eq.~(\ref{eq:texact}), so that
\begin{equation}
|t(E)|^2 \simeq \frac{16(E_F-E)(V_0-E_F)}{(V_0-E)^2} e^{-2L\sqrt{2m^*(V_0-E_F)/\hbar^2}} ,
\end{equation}
where we have substituted $V_0\rightarrow V_0-E$ and $E_F\rightarrow E_F-E$, as appropriate for the Q1D model, and we adopt the notation $L=2b$ used in the main text.

We now compute the density of subband levels, $\partial N_w/\partial E$.
The subbands are solutions to the Schr\"{o}dinger equation for a finite square well of width $w$ and depth $V_0$.
Our goal is to determine the number ($N_w$) of energy eigenstates ($E$) that lie below a fixed energy level ($E_F$) in the square well.
If the square well is centered at position $x=0$, then the symmetry of the eigenfunctions is either even or odd.
In the first case, the wave function is given by
\begin{gather}
\psi_M(x) = A_M \cos (kx) \quad\quad (|x|\leq w/2) , \\
\psi_R(x) = A_R \, e^{-qx} \quad\quad (x>w/2) ,
\end{gather}
while in the second case,
\begin{gather}
\psi_M(x) = A_M \sin (kx) \quad\quad (|x|\leq w/2) , \\
\psi_R(x) = A_R \, e^{-qx} \quad\quad (x>w/2) .
\end{gather}
For the energy eigenvalue $E$ and barrier height $V_0$, we have 
\begin{equation}
E= \frac{\hbar^2 k^2}{2m^*} \quad\quad \text{and} \quad\quad
V_0-E=\frac{\hbar^2 q^2}{2m^*} .
\end{equation}
By equating the wave function and its derivative at the edge of the well, we obtain the eigenvalue conditions
\begin{gather}
\tan(kw/2) =\left( \frac{2m^*V_0}{\hbar^2k^2}-1\right)^{1/2} \quad\quad (\text{even case}) , \label{eq:even} \\
\tan(kw/2) =-\left( \frac{2m^*V_0}{\hbar^2k^2}-1\right)^{-1/2} \quad (\text{odd case}) .  \label{eq:odd}
\end{gather}

The transcendental solutions to Eqs.~(\ref{eq:even}) and (\ref{eq:odd}) can be visualized by plotting their left- and right-hand-sides on the same graph, as functions of $k$.
The solutions are obtained at the intersections of the curves, with each intersection representing a subband.
In general, Eq.~(\ref{eq:even}) has at least one real solution.
However, when $E_F$ is small, there may be no physical solutions (i.e., no filled subbands).

The total number of solutions to the eigenvalue problem specified in Eqs.~(\ref{eq:even}) and (\ref{eq:odd}) is closely related to the number of half-oscillations of $\tan (kw/2)$ that occur in the range $0<k\leq k_F$, where $k_F$ is the Fermi wave vector.
Graphical inspection shows that this number is given by 
\begin{equation}
N_w=\left\lfloor \frac{w}{\pi} \sqrt{\frac{2 m^* E_F}{\hbar^2}} \right\rfloor +\theta (E_F,V_0,w) , 
\label{eq:Nw}
\end{equation} 
where $\lfloor x \rfloor$ is the integer floor function, and $\theta (x)$ is a Heaviside step function, which takes the value 0 or 1, depending on the values of its arguments.
For large $w$, the Heaviside term is much smaller than the first term in Eq.~(\ref{eq:Nw}), and we can approximate 
\begin{equation}
N_w\simeq \frac{w}{\pi} \sqrt{\frac{2m^*E_F}{\hbar^2}} =\frac{k_Fw}{\pi} . 
\end{equation}
The latter result is also known in the context of subband filling in a quantum point contact.\cite{DattaBook}
In the analysis leading to Eq.~(\ref{eq:Nw}), the subbands occur at nearly equally spaced $k$ values, with the density $\partial N_w/\partial k = w/\pi$.  
Hence we obtain 
\begin{equation}
\frac{\partial N_w}{\partial E} = \frac{w}{\pi} \sqrt{\frac{m^*}{2\hbar^2 E}} ,
\end{equation}
for the subband energy $E$.

Performing the integral in Eq.~(\ref{eq:GNwint}), we finally obtain 
\begin{widetext}
\begin{equation}
\frac{G}{w}\simeq \left(\frac{4e^2}{h}\right) \frac{8(V_0-E_F)}{\pi V_0} \sqrt{\frac{2m^*E_F}{\hbar^2}} 
\left[ \left( \frac{V_0+E_F}{2\sqrt{V_0E_F}} \right)
\ln \left(\frac{\sqrt{V_0}+\sqrt{E_F}}{\sqrt{V_0}-\sqrt{E_F}} \right) -1 \right]
e^{-2L\sqrt{2m^*(V_0-E_F)/\hbar^2}} . \label{eq:Gasymp}
\end{equation}
\end{widetext}
This expression represents the tunnel conductance between two wide leads in a 2DEG, in the Q1D model.
It is plotted with solid lines in Figs.~3(b) and (d) of the main text.
When the lead is wide and the tunnel barrier is short, Eq.~(\ref{eq:Gasymp}) provides a reasonable estimate for the conductance in a 2D geometry.
When the barrier is long, transverse spreading of the wave function becomes important inside the tunnel barrier, and the 2D conductance is strongly suppressed compared to Eq.~(\ref{eq:Gasymp}).

\end{document}